\documentstyle[aps,multicol,prb]{revtex}

\input{epsf}
\epsfverbosetrue
\epsfclipon

\newcommand{\putfigxsz}[4]{
   \begin{figure}\begin{center}\mbox{\epsfxsize #4
   \epsffile{#1}}
   \end{center}
   \end{figure}
   }

\begin{document}
\draft
\title{Weak localisation, hole-hole interactions and the\\
 ``metal"-insulator transition in two dimensions}
\author{M.Y. Simmons\cite{Now-at}, A.R. Hamilton\cite{Now-at}, M. Pepper, E.H. Linfield,
 P.D. Rose, and D.A. Ritchie}
\address{Cavendish Laboratory, University of Cambridge, Madingley Road, Cambridge CB3 OHE,
 U.K.}

\date{20th October 1999}
\maketitle
\begin{abstract}
A detailed investigation of the metallic behaviour in high quality
GaAs-AlGaAs two dimensional hole systems reveals the presence of
quantum corrections to the resistivity at low temperatures.
Despite the low density ($r_{s}>10$) and high quality of these
systems, both weak localisation (observed via negative
magnetoresistance) and weak hole-hole interactions (giving a
correction to the Hall constant) are present in the so-called
metallic phase where the resistivity decreases with decreasing
temperature. The results suggest that even at high $r_{s}$ there
is no metallic phase at T=0 in two dimensions.

\end{abstract}


\begin{multicols}{2}
Since the claimed observation of metallic behaviour in strongly
interacting two-dimensional (2D) systems over 5 years
ago~\cite{Kravchenko} experimentalists have tried to provide data
from which an understanding of the conduction processes in high
quality 2D systems can be obtained. Initial studies of these new
strongly interacting systems revealed that the resistivity data
can be ``scaled" over a wide range of temperatures indicating the
presence of a true phase transition between insulating and
metallic states~\cite{Kravchenko,Popovic-Coleridge,Simmons}.
Following this an empirical formula for $\rho$(T) has been put
forward which fits all the available experimental data of the
metallic state~\cite{Pudalov97,Hanein}. This formula describes a
saturation of the resistivity as the temperature is reduced,
giving a finite resistance at T=0, further testifying the
existence of a 2D metallic state. Despite these results the nature
of the metallic state and whether it really persists to the zero
of temperature remains unclear. Early
theoretical~\cite{Abrahams,Altshuler} and
experimental~\cite{Uren-BDT} studies of weakly interacting systems
(low $r_{s}$) revealed that the presence of any disorder would
give rise to logarithmic corrections to the conductivity. Since
these corrections become increasingly important as $T$ is reduced
the question of what happens to the metallic behaviour as
T$\rightarrow 0$ in 2D systems remains.

This paper reports the observation of both weak localisation and
weak hole-hole interactions in the ``metallic" phase of a high
quality 2D GaAs hole system. First we demonstrate that the system
studied here exhibits all of the characteristics previously
associated with the 2D ``metal"-insulator transition.
Magnetoresistance measurements are then used to extract the
logarithmic corrections to the Drude conductivity at low
temperatures. The data show that: (1) the anomalous exponential
decrease of resistivity with decreasing temperature in the
metallic phase is not due to quantum interference or strong
interaction effects, (2) phase coherence is preserved in the
metallic regime with evidence for normal Fermi liquid behaviour,
and (3) hole-hole interactions provide a localising correction to
the conductivity.

The sample used here is a gated, modulation doped GaAs quantum
well grown on a (311)A substrate~\cite{Simmons}. Four terminal
magnetoresistance measurements were performed at temperatures down
to 100mK using low frequency (4~Hz) ac lockin techniques and
currents of 0.1-5~nA to avoid electron heating. The hole density
could be varied in the range
$0-3.5{\times}10^{11}~\text{cm}^{\text{-2}}$, with a peak mobility
of $2.5{\times}10^{5}
~\text{cm}^{\text{2}}\text{V}^{\text{-1}}\text{s}^{\text{-1}}$.
Only the heavy hole ($|M_J|$=3/2) subband is occupied, although
there is some mixing between light and heavy hole bands for
$|k|>0$.

Figure~\ref{fig1}(a) shows the temperature dependence of the $B$=0
resistivity $\rho$ plotted for carrier densities close to the
transition, from
$p_{s}=3.2-5.6{\times}10^{10}~\text{cm}^{\text{-2}}$. Strongly
localised behaviour is observed at the lowest carrier densities,
with $\rho$ taking the familiar form for variable range hopping:
$\rho(T)=\rho_{\text{VRH}} \exp[(T/T_{\text{VRH}})^{-m}]$, with
m=1/2 far from the transition and m=1/3 close to the transition. A
transition from insulating to metallic behaviour occurs as the
carrier density is increased, with a critical density of
$p_{c}=4.6{\times}10^{10}~\text{cm}^{\text{-2}}$ ($r_{s}=12$) at
the transition.  Above this critical density the resistivity drops
markedly as the temperature is reduced, although it is difficult
to see this drop on the logarithmic axis of Fig. 1(a).

The metallic behaviour can be seen more clearly in the ``scaled"
data shown in Fig. 1(b). Each $\rho$(T) trace was individually
scaled along the T-axis in order to collapse all the data onto one
of two separate branches. It has been suggested that the ability
to scale the data both in the strongly localised and metallic
branches is evidence for a phase transition between insulating and
metallic states in a 2D system\cite{Kravchenko}. More recently
Pudalov {\it et al.}~\cite{Pudalov97} have shown that $\rho(T)$ in
the metallic regime is well described by:
\begin{equation}
\rho(T)=\rho_0 + \rho_1 \exp[-T_a/T]
\end{equation}

Fig. 1(c-e) shows the temperature dependent resistivity for three
different carrier densities in the metallic regime, with  fits to
Eqn.~(1) shown as dashed lines. In Fig. 1(c), at a density close
to the transition, saturation of the resistivity is just visible
at the lowest temperatures (100mK). As the density is increased
and we move further into the metallic regime this saturation
becomes visible at higher temperatures until at the highest
density $\rho(T)$ saturates below 350mK. The empirical formula
(1), which characterises the metallic behaviour observed in all 2D
systems therefore dictates a saturation of $\rho(T)$ as $T
\rightarrow 0$. Although different from the scaling analysis of
Ref. [1] and shown in Fig. 1(b), it is still consistent with the
existence of a 2D metal-insulator transition because $\rho(T)$
remains finite as $T \rightarrow 0$. Early studies of weakly
interacting, disordered 2D systems ($r_{s}\sim 4$)~\cite{Uren-BDT}
demonstrated that both weak localisation and weak
electron-electron interactions caused a logarithmic reduction of
the conductivity as $T \rightarrow 0$. More recently it has been
shown that the same interaction effects occur in slightly less
disordered samples ($r_{s}\sim 6$) that exhibit ``metallic
behaviour", at high carrier densities, far from the
metal-insulator transition\cite{Hamilton}. However neither the
scaling analysis in Fig. 1(b) nor the empirical Eqn.~(1) address
what has happened to these logarithmic corrections near the
metal-insulator transition, and whether the conductivity remains
finite as $T\rightarrow0$.

We now turn to one of two main results of this paper.  Figure 2
shows the temperature dependence of the $B$=0 resistivity (left
hand panel) and magnetoresistance (right hand panel) at different
densities on both sides of the ``metal"-insulator transition. In
Fig. 2(a) we are just on the insulating side of the transition.
The left hand panel shows that $\rho(T)$ is essentially
$T$-independent down to 300mK and then increases by 2.5$\%$ as the
temperature is further reduced. This weak increase in the
resistivity has been previously taken as evidence for weak
localisation and weak electron-electron interaction
effects\cite{Hamilton,Pudalov}. It is however not possible to
determine the precise origins of this weak increase in resistivity
solely from the $B$=0 data, and we therefore look at the
magnetoresistance shown in the right hand panel of Fig. 2(a). A
characteristic signature of weak localisation is a strong
temperature dependent negative magnetoresistance, since the
perpendicular magnetic field breaks time reversal symmetry,
removing the phase coherent back-scattering. As observed
previously there is no evidence of weak localisation for
temperatures down to 300mK in these high quality
samples~\cite{Simmons}. However, as $T$ is lowered below 300mK a
strong negative magnetoresistance peak develops as phase coherent
effects become important, mirroring the small increase in the
resistivity at $B$=0.

Increasing the carrier density brings us into the metallic regime
(Fig. 2(b)) where the exponential drop in the resistivity with
decreasing temperature predicted from Eqn. (1) starts to become
visible. The upturn in $\rho(T)$ marked by the arrow has moved to
lower temperatures and the negative magnetoresistance in the right
hand panel has become less pronounced. Further increasing the
density (Fig. 2(c)) causes the metallic behaviour to become
stronger, with the upturn in $\rho(T)$ moving to even lower
temperatures, until at
$p_{s}=5{\times}10^{10}~\text{cm}^{\text{-2}}$ the upturn is no
longer visible within the accessible temperature measurement
range. However, the magnetoresistance still exhibits remnants of
the weak localisation temperature dependent peak at $B$=0. The
weak localisation is therefore always present and is neither
destroyed in the metallic regime, nor is it ``swamped" by the
exponential decrease in resistivity with decreasing temperature.
Instead what can clearly be seen in the left hand panel of Fig. 3,
is that the upturn in $\rho(T)$ due to weak localisation marked by
the arrows, moves to lower $T$ as the carrier density is
increased. This is not surprising since as we move further into
the metallic regime both the conductivity and therefore the mean
free path ($l\propto\sigma/\sqrt{p_{s}}$) increase, such that the
weak localisation corrections are only visible at lower
temperatures (larger $l_{\phi}$).

 In contrast to
experimental\cite{Pedersen} studies of high carrier density hole
gas quantum well samples there are no signs of weak
anti-localisation in these low density samples. This is perhaps to
be expected since recent theoretical work\cite{Averkiev} has
predicted that the magnetoresistance behaviour is determined by
the degree of heavy-hole/light-hole mixing at the Fermi energy,
which is characterised by the parameter $k_{F}a/\pi$, where $a$ is
the width of the quantum well. In our sample the carrier
concentration is small, such that $k_{F}a/\pi\ll 1$, and only
negative magnetoresistance is expected.

In Fig. 3 (a) we fit the temperature dependent magnetoconductance
data to the formula of Hikami {\it et al.}\cite{Hikami}.

\begin{equation}
\Delta\sigma(B) = \frac{- e^{2}}{\pi h}
\left[\Psi\left(\frac{1}{2} + \frac{B_{\phi}}{B}\right) -
\Psi\left(\frac{1}{2} + \frac{B_{0}}{B}\right)\right]
\end{equation}

where $\Psi(x)$ is the digamma function, B$_{0}$ and B$_{\phi}$
are characteristic magnetic fields related to the transport
scattering rate and the phase relaxation rate.  We obtain
$\sigma_{xx}$ by matrix inversion of $\rho_{xx}$ and $\rho_{xy}$.
Using Eqn. (2) we fit the experimental data just on the metallic
side of the transition
($p_{s}=4.7{\times}10^{10}~\text{cm}^{\text{-2}}$) for different
temperatures as shown in Fig. 3(a). These fits are in good
agreement with the experimental data, and from this it is possible
to extract the fitting parameter B$_{\phi}$ and thus the phase
relaxation time $\tau_{\phi}$.

Figure 3(b) shows the temperature dependence of the phase breaking
rate, $1/\tau_{\phi}$ for three different densities on both sides
and close to the ``metal"-insulator transition. The phase breaking
rate falls approximately linearly with decreasing temperature for
all three traces. The linear dependence agrees well with that
predicted for disorder enhanced hole-hole
scattering~\cite{Altshuler82}, where $1/\tau_{\phi}\sim
2k_{B}T/(\hbar k_{F}l)$. This phase breaking mechanism should only
depends on $ k_{F}l$ and not on the carrier density, mobility or
interaction strength. It is therefore particularly noteworthy that
the phase breaking rates in these low density p-GaAs samples, with
$2.5<k_{F}l<5$, are almost identical to those found in n-type
silicon MOSFETs~\cite{Davies} with $k_{F}l\sim 1$, despite a
factor of 20 difference in the carrier densities (see data in Fig.
3(b)). This agreement with scattering limited electron lifetime
suggests that the electron states are only mildly perturbed by the
strong interactions and essentially remain Fermi liquid like.

Another important feature of these results is that there is little
variation in $\tau_{\phi}$ with density and in particular there is
no dramatic change in $\tau_{\phi}$ as we cross from insulating
($p_{s}= 4.5{\times}10^{10}~\text{cm}^{\text{-2}}$) to metallic
behaviour ($p_{s}= 5.2{\times}10^{10}~\text{cm}^{\text{-2}}$).
There is therefore no reflection of the exponential decrease of
$\rho(T)$ with decreasing temperature in the phase breaking rate.
This implies that whatever mechanism is causing metallic behaviour
does not suppress weak localisation as originally believed and is
further evidence that the system is behaving as a Fermi liquid.
Since all models of the resistivity in the metallic
phase~\cite{Pudalov97,Mills} predict that the exponential drop
saturates at low temperatures, our data shows that localisation
effects will again take over as T$\rightarrow$0.

Finally we address the role of electron-electron (hole-hole)
interactions in the 2D metallic phase - the second important
result from this paper. Unlike weak localisation, interactions not
only affect the $B=0$ resistivity, but also cause a correction to
the Hall resistance:

\begin{equation}
\frac{\Delta R_{H}} {R_{H}} = -2\frac{\Delta\sigma_{I}}{\sigma}
\label{EQHall}
\end{equation}

By measuring the low field Hall effect it is thus possible to
distinguish between weak localisation and interaction
effects\cite{Uren2}. Figure 4(a) shows the Hall resistivity
$\rho_{xy}$ measured on the metallic side of the transition (i.e.
where the zero field resistivity shows an exponential drop with
decreasing temperature as shown in Fig. 2(c)). The data reveal a
small, but significant, decrease of the Hall slope with increasing
temperature. Whilst a series of different temperatures traces from
100-700mK were taken only three of these traces are presented for
clarity. Upon closer investigation this small decrease of the Hall
slope is found to vary as $\log(T)$. We extract the interaction
correction to the zero field conductivity, $\Delta\sigma_{I}$ from
the temperature dependent Hall data using equation (3). Figure
4(b) shows a plot of the interaction correction for different
carrier densities on both sides of the transition. All the data
collapses onto a single line, clearly demonstrating a $\log(T)$
dependence of $\Delta\sigma_{I}$, which reduces the conductivity
to zero as T$\rightarrow$0.

Logarithmic corrections to the Hall resistivity have previously
been observed in studies of interaction effects in high density
electron systems~\cite{Uren2}. It is perhaps surprising that
results observed in, and derived from, weakly interacting systems
apply to our system where interactions are strong and $r_{s}>10$.
Nevertheless, we find reasonable agreement between the magnitude
of the logarithmic corrections due to interactions in our system
and those predicted by Altshuler et al.\cite{Altshuler} (within a
factor of 2). As with the phase coherent effects this logarithmic
correction due to hole-hole interactions is independent of whether
we are in the insulating or metallic phase and is present despite
the exponential drop in resistivity. This is the first proof that
electron-electron interactions are not responsible for the 2D
``metal"-insulator transition observed in high mobility (low
$E_{F}$) systems.

In summary we have presented a comprehensive study of localisation
and interaction effects in a high mobility two-dimensional hole
gas sample that shows all the signatures of a $B=0$
``metal"-insulator transition. The results clearly demonstrate
that neither phase coherent effects nor electron-electron
interactions are responsible for the apparent 2D metal-insulator
transition. Both of these effects are present in the metallic
regime and both give rise to localising corrections to the
conductivity at low temperatures. The importance of phase coherent
effects in studies of the metallic behaviour in high mobility 2D
systems has not previously been recognised because the mean free
path  in these systems is large, so that weak localisation is only
observable at very low, often inaccessible, temperatures.
Nevertheless we clearly observe negative magnetoresistance
characteristic of weak localisation in the so-called metallic
phase. There is no suppression of phase coherent effects in the
metallic regime. Instead the weak localisation simply moves to
lower temperatures as we go further into the metallic regime. Most
importantly we demonstrate that the metallic behaviour is not due
to hole-hole interactions, because these also cause a logarithmic
localising correction to the conductivity. These results strongly
suggest that the metallic behaviour must be a finite temperature
effect, and that as T$\rightarrow$0 the old results of scaling
theory and weak electron-electron interactions remain valid -
there is no genuine 2D metallic phase.

We are indebted to D.E. Khmel'nitskii, D. Neilson, and B.
Altshuler for many useful discussions. This work was funded by
EPSRC (U.K.). MYS acknowledges a QEII Fellowship from the
Australian Research Council.\\

\begin{minipage}{8.5cm}
\noindent
\begin{figure}[tbph]
\putfigxsz{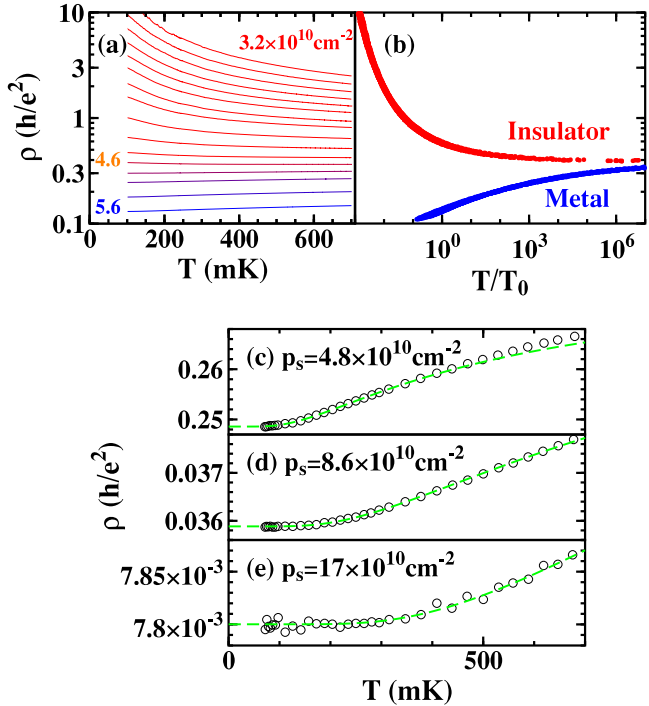}{lab}{cap}{8.5cm} \caption{(a) Temperature
dependence of the resistivity at densities from
$p_{s}=3.2-5.6{\times}10^{10}~\text{cm}^{\text{-2}}$. (b) The same
data, scaled to collapse onto metallic and insulating branches.
(c) Temperature dependence of the resistivity in the metallic
regime at carrier densities of 4.8, (d) 8.6 and (e)
17${\times}10^{10}~\text{cm}^{\text{-2}}$. The dashed lines show
the fit to Eqn. (1). } \label{fig1} \vspace{0.1in}
\end{figure}
\end{minipage}

\begin{minipage}{8.5cm}
\noindent
\begin{figure}[tbph]
\putfigxsz{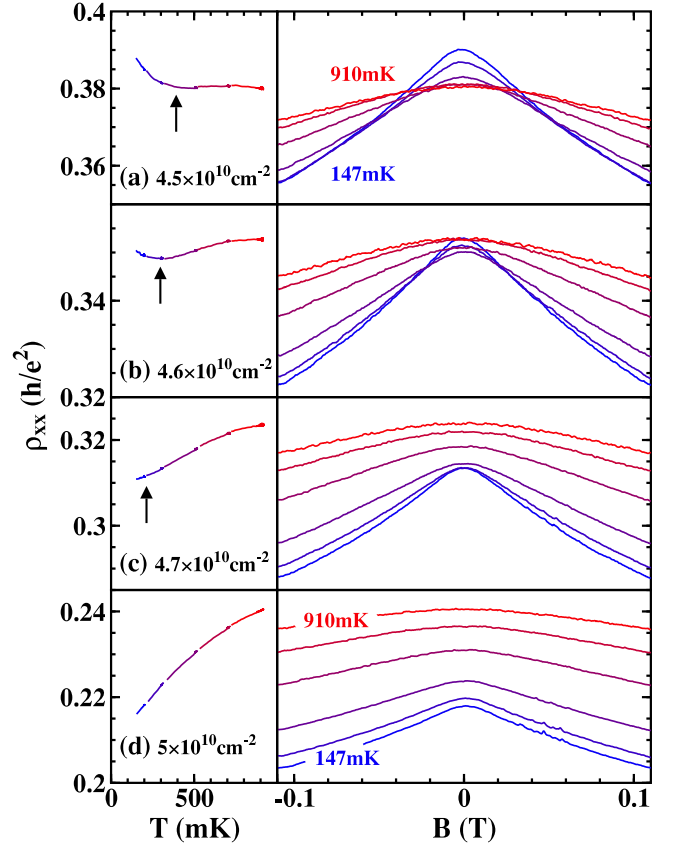}{lab}{cap}{8.5cm} \caption{(a-d) The left hand
panels show resistivity at B=0 versus temperature data,
illustrating the transition from insulating to metallic behaviour
as the density increases. The right hand panels show the
corresponding magnetoresistance traces for temperatures of 147,
200, 303, 510, 705 and 910mK.} \label{fig2}
\end{figure}
\end{minipage}

\begin{minipage}{8.5cm}
\noindent
\begin{figure}[tbph]
\putfigxsz{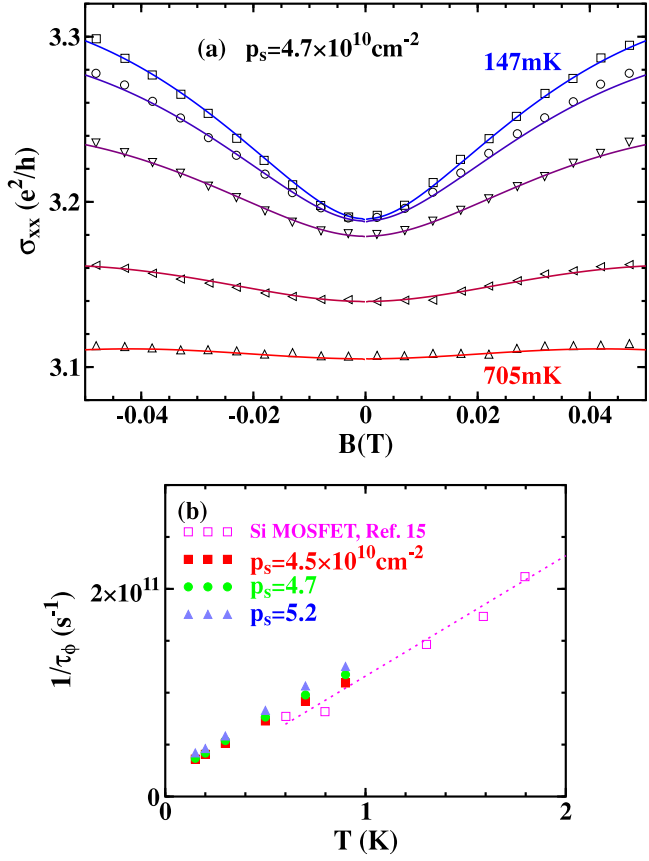}{lab}{cap}{8.5cm} \caption{(a) The
magnetoconductivity, $\sigma_{xx}$ just on the metallic side of
the transition for temperatures of 147, 200, 303, 510 and 705~mK.
(b) A plot of $1/\tau_{\phi}$ versus temperature for densities
close to the ``metal"-insulator transition. Solid symbols are data
obtained from this study; open symbols are data from Si MOSFETs,
Ref. [15]} \label{fig3}
\end{figure}
\end{minipage}

\begin{minipage}{8.5cm}
\noindent
\begin{figure}[tbph]
\putfigxsz{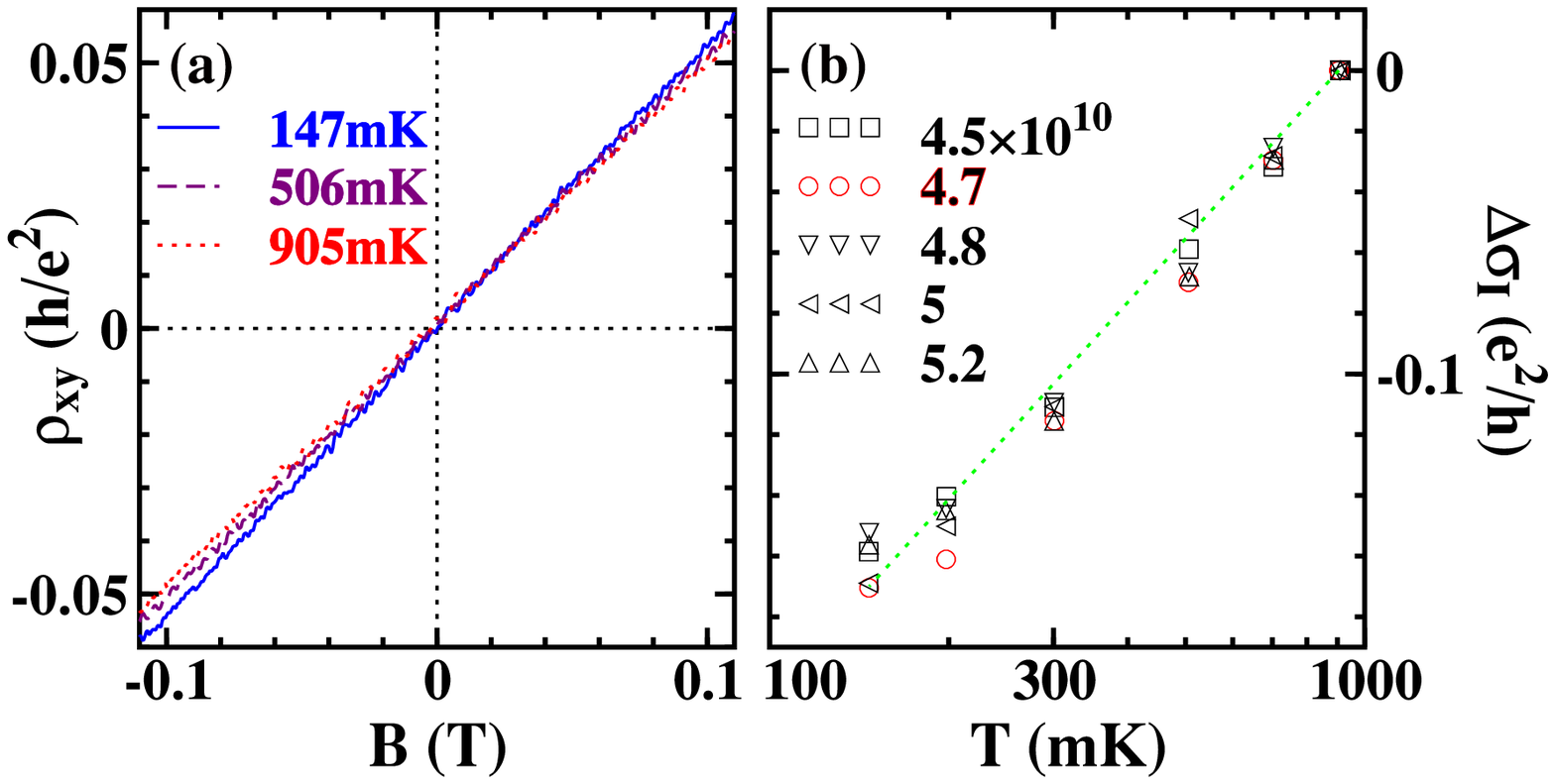}{lab}{cap}{8.5cm} \caption{(a) Hall
resistivity for different temperatures at a density of
4.7${\times}10^{10}~\text{cm}^{\text{-2}}$ (b) the logarithmic
correction to the conductivity for densities close to the
transition.} \label{fig4}
\end{figure}
\end{minipage}

\end{multicols}

\begin{references}

\bibitem[*]{Now-at}Present address: Semiconductor
Nanofabrication Facility, University of New South Wales, Sydney
2052, Australia

\bibitem{Kravchenko}S.V. Kravchenko {\it et al.},
Phys.  Rev.  B {\bf 50}, 8039 (1994); S.V. Kravchenko {\it et
al.}, Phys. Rev. B {\bf 51}, 7038 (1995).

\bibitem{Popovic-Coleridge}D. Popovi\'{c}, A.B. Fowler and S. Washburn,
Phys. Rev. Lett {\bf 79}, 1543 (1997);P.T. Coleridge {\it et al.},
Phys. Rev. B {\bf 56}, R12764 (1997).

\bibitem{Simmons}M.Y. Simmons {\it et al.},  Phys. Rev. Lett. {\bf 80},
1292 (1998).

\bibitem{Pudalov97}V.M. Pudalov {\it et al.}, JETP
Lett. {\bf 66} 175 (1997).

\bibitem{Hanein}Y. Hanein {\it et al.}, Phys. Rev. Lett. {\bf 80},
1288 (1998).

\bibitem{Abrahams}E. Abrahams {\it et al.}, Phys. Rev. Lett. {\bf 42},
690 (1979); L.P. Gor'kov, A.I. Larkin and D.E. Khmel'nitskii, JETP
Lett. {\bf 30}, 229 (1979).

\bibitem{Altshuler}B.L. Altshuler, A.G. Aronov, and P.A. Lee, Phys. Rev.
Lett. {\bf 44} 1288 (1980); B.L. Altshuler and A.G. Aronov, in
\emph{electron-electron interactions in disordered systems}, Efros
and Pollak, (1985).

\bibitem{Uren-BDT}M.J. Uren {\it et al.}, J. Phys. C{\bf 13}, L985 (1980);
D.J. Bishop, D.C. Tsui, and R.C. Dynes, Phys. Rev. Lett {\bf 44},
1153 (1980).

\bibitem{Hamilton}A.R. Hamilton {\it et al.}, Phys. Rev. Lett. {\bf 82}, 1542 (1999).

\bibitem{Pudalov}V.M. Pudalov {\it et al.} , JETP Lett. {\bf 68}, 415 (1998).

\bibitem{Pedersen}S. Pedersen {\it et al.}, Phys. Rev. B {\bf
60}, 4880 (1999).

\bibitem{Averkiev}N.S. Akerkiev, L.E. Golub and G.E. Pikus, JETP
{\bf 86}, 780 (1998).

\bibitem{Hikami}S. Hikami, A.I. Larkin, Y. Nagaoka, Prog. Theor.
Phys. {\bf 63}, 707 (1980).

\bibitem{Altshuler82}B.L. Altshuler, A.G. Aronov and D.E.
Khmel'nitskii, J. Phys. C {\bf 15}, 7367 (1982).

\bibitem{Davies}R.A. Davies and M. Pepper, J. Phys. C {\bf 16},
L353 (1983).

\bibitem{Mills}A.P. Mills {\it et al.}, Phys. Rev. Lett. {\bf 83}, 2805 (1999).

\bibitem{Uren2}M.J. Uren {\it et al.}, J. Phys. C
{\bf 14}, 5737 (1981).

\end{references}
\end{document}